\documentstyle[12pt]{article}
\begin{document}

\begin{titlepage}

\begin{flushright} 
EPHOU-97-003

February, 1997
\end{flushright}

\vspace{15mm}
\begin{center} 

{\Large Quantization of Infinitely Reducible \\}
{\Large Generalized Chern-Simons Actions in Two Dimensions}\\
\vspace{1cm}
{\bf {\sc Noboru Kawamoto, Kazuhiko Suehiro, Takuya Tsukioka\\
and Hiroshi Umetsu}}\\
{\it{ Department of Physics, Hokkaido University }}\\
{\it{ Sapporo, 060, Japan}}\\
{kawamoto, suehiro, tsukioka, umetsu@phys.hokudai.ac.jp}
\end{center}
\vspace{2cm}

\begin{abstract}
We investigate the quantization of two-dimensional version 
of the generalized Chern-Simons actions which were proposed previously.
The models turn out to be infinitely reducible and thus we need 
infinite number of ghosts, antighosts and the corresponding antifields.
The quantized minimal actions which satisfy the master equation of 
Batalin and Vilkovisky have the same Chern-Simons form.
The infinite fields and antifields are successfully controlled 
by the unified treatment of generalized fields with quaternion algebra.
This is a universal feature of generalized Chern-Simons theory 
and thus the quantization procedure can be naturally extended to 
arbitrary even dimensions. 
\end{abstract}

\end{titlepage}

\renewcommand{\theequation}{\arabic {section}.\arabic{equation}}

\setlength{\baselineskip}{7mm}

%%%%%%%%%%%%%%%%%%%%%%%%%%%%%%%%%%%%%%%%%%%%%

\section{Introduction} 
\setcounter{equation}{0}

The Chern-Simons action has many applications 
for physical mechanisms and formalisms. 
In particular it was used to formulate 
three-dimensional Einstein gravity~\cite{w1}.   
Two of possible reasons why three-dimensional Einstein gravity was 
successfully formulated by the Chern-Simons action are based on the facts 
that the action is formulated by differential forms on the one hand and 
the three-dimensional Einstein gravity has no dynamical degrees of freedom 
on the other hand.     

One of the authors (N.K.) and Watabiki have proposed 
a new type of topological actions in arbitrary dimensions 
which have the Chern-Simons form~\cite{kw1,kw2,kw3}.
The actions have the same algebraic structure as the ordinary Chern-Simons 
action and are formulated by differential forms. 
It was shown that two-dimensional topological gravities~\cite{kw2}
and a four-dimensional topological conformal gravity~\cite{kw3}
were formulated by the even-dimensional version 
of the generalized Chern-Simons actions.

It is interesting to ask if the models defined by the generalized Chern-Simons 
actions are well-defined in the quantum level 
and thus lead to the quantization of topological gravity. 
It turns out that 
the quantization of the generalized Chern-Simons action is highly nontrivial.
The reasons are two folds: Firstly the action has a zero form square term 
multiplied by the highest form and thus breaks regularity condition. 
Secondly the theory is highly reducible, in fact infinitely reducible, 
as we show in this paper. 
Thus the models formulated by the generalized Chern-Simons actions 
provide its own interesting problems for the known quantization 
procedures such as Batalin and Vilkovisky formulation 
of the master equation~\cite{bv}, 
Batalin, Fradkin and Vilkovisky Hamiltonian formulation~\cite{bfv}
and the quantization procedure of cohomological perturbation~\cite{ht}. 

It was shown in the quantization of the simplest abelian version of 
generalized Chern-Simons action that the particular type of regularity 
violation does not cause serious problems for the quantization~\cite{kos}. 
In this paper we investigate nonabelian version of Chern-Simons actions 
which turn out to be infinitely reducible. 
We show that the quantization of this infinitely reducible system 
can be treated successfully by the unified treatment of fields and 
antifields of the generalized Chern-Simons theory.
It is interesting to note that the nonabelian version of the generalized 
Chern-Simons actions provide the most fruitful examples for the quantization 
of infinitely reducible systems among the known examples such as 
Brink-Schwarz superparticle~\cite{bs}, 
Green-Schwarz superstring~\cite{gs} 
and covariant string field theories~\cite{w2}. 

%%%%%%%%%%%%%%%%%%%%%%%%%%%%%%%%%%%%%%%%%%

\section{Generalized Chern-Simons theory} 
\setcounter{equation}{0}

The generalized Chern-Simons theory 
is a generalization of the ordinary three dimensional Chern-Simons theory 
into arbitrary dimensions~\cite{kw1,kw2,kw3}.
The main point of the generalization is 
to extend a one form gauge field to a quaternion valued
generalized gauge field ${\cal A}$ which contains forms 
of all possible degrees.
Correspondingly a gauge symmetry is extended 
and it is described by a quaternion valued gauge parameter ${\cal V}$.
It was shown that this formulation can naturally incorporate fermionic
gauge fields and parameters as well. 
In the most general form, a generalized gauge field ${\cal A}$ 
and a gauge parameter ${\cal V}$ are defined by the following component form,
\begin{eqnarray}
 {\cal A} & = & {\mbox{\bf 1}}\psi + {\mbox{\bf i}} \hat{\psi} +  
            {\mbox{\bf j}} A + {\mbox{\bf k}} \hat{A}, \label{eqn:gc} \\
 {\cal V} & = & {\mbox{\bf 1}} \hat{a} + {\mbox{\bf i}} a +  
            {\mbox{\bf j}} \hat{\alpha} + {\mbox{\bf k}} \alpha, 
                                                      \label{eqn:gcc}
\end{eqnarray}
where $( \psi, \alpha )$, $( \hat{\psi}, \hat{\alpha} )$, 
$( A, a )$ and  $( \hat{A},\hat{a} )$ are direct sums of 
fermionic odd forms, fermionic even forms, bosonic odd forms 
and bosonic even forms, respectively, and they take values on a gauge algebra.
The bold face symbols ${\bf 1}$, ${\bf i}$,  ${\bf j}$ and ${\bf k}$ 
are elements of quaternion.
The two types of component expansions (\ref{eqn:gc}) and (\ref{eqn:gcc}), 
which belong to $\Lambda_-$ and $\Lambda_+$ classes, 
can be regarded as generalizations of odd forms and even forms, respectively. 
In the case of even-dimensional formulation 
a gauge algebra can simply be chosen 
as such an algebra as is closed within commutators and anticommutators.
In this case the elements in $\Lambda_-$ and $\Lambda_+$ classes fulfill 
the following $Z_2$ grading structure;
\begin{eqnarray}
   \lambda_+\lambda_+ \in \Lambda_+, \ \ \ \ 
   \lambda_-\lambda_+ \in \Lambda_-, \ \ \ \ 
   \lambda_-\lambda_- \in \Lambda_+,         \label{eqn:z2} 
\end{eqnarray}
where $\lambda_+ \in \Lambda_+,~ \lambda_-\in \Lambda_-$. 
In general, 
graded Lie algebra is necessary to accommodate odd-dimensional formulation.

The even-dimensional version of actions proposed by Kawamoto and Watabiki 
possess the following Chern-Simons form~\cite{kw1},
\begin{eqnarray}
 S = \frac {1}{2} \int _M \ {\mbox{Tr}}_{\mbox{\bf {k}}} 
             \left( {\cal A}Q{\cal A} + \frac {2}{3} {\cal A}^3 \right),
                                                             \label{eqn:ga}
\end{eqnarray}
where 
$Q  =  {\mbox{\bf j}}d \in \Lambda_-$ 
is the exterior derivative and 
${\mbox{Tr}}_{\mbox{\bf {k}}} \left( \cdots \right) $ 
is defined so as to pick up only the coefficient of 
${\bf k}$ from $\left( \cdots \right)$ 
and take the trace of the gauge algebra. 
The ${\bf k}$ component of an element in the $\Lambda_-$ class includes 
only bosonic even forms and thus the action (\ref{eqn:ga}) 
leads to an even-dimensional one.
We then need to pick up $d$-form terms 
corresponding to the $d$-dimensional manifold $M$.
Since this action has the same structure 
as the ordinary three-dimensional Chern-Simons action, 
it is invariant under the following gauge transformation, 
\begin{eqnarray}
 \delta {\cal A} = [ \ Q+{\cal A} \ , \ {\cal V} \ ].\label{eqn:gt}
\end{eqnarray}
It should be noted that this symmetry is much larger than
the usual gauge symmetry since the gauge parameter ${\cal V}$ contains
many parameters of various forms.
Since anticommutators as well as commutators for elements of the gauge algebra 
appear in the explicit form of the gauge transformations, 
we need to use an algebra which is closed 
within commutators and anticommutators.
A specific example of the algebra is realized by Clifford algebra.
In general a generalized gauge theory can be formulated 
for a graded Lie algebra which includes supersymmetry algebra 
as a special example~\cite{kw1}.

The equation of motion of this theory is 
\begin{eqnarray}
 {\cal F} = 0, \label{eqn:ge}
\end{eqnarray}
where ${\cal F}$ is a generalized curvature, given by
\begin{eqnarray}
 {\cal F} = ( Q + {\cal A} )^2 = Q{\cal A} + {\cal A}^2. \label{eqn:gf}
\end{eqnarray}
%

%%%%%%%%%%%%%%%%%%%%%%%%%%%%%%%%%%%%%%%%%%%

\section{Infinite reducibility of two-dimensional models} 
\setcounter{equation}{0}

Hereafter we consider the action (\ref{eqn:ga}) in two dimensions 
with a nonabelian gauge algebra 
as a concrete example although we will see that models 
in arbitrary even dimensions can be treated in the similar way.
A simple example for nonabelian gauge algebras 
is given by Clifford algebra $c(0,3)$ generated 
by $\{T^a\}=\{1,i\sigma^k; k=1,2,3\}$ where $\sigma^k$'s are 
Pauli matrices~\cite{kw2}.
For simplicity we omit fermionic gauge fields and parameters 
in the starting action and gauge transformations.
It is, however, easy to recover them in the subsequent formulation. 
Then the action expanded into components is given by 
\begin{eqnarray}
 S_0 = - \int \ d^2 x {\mbox{Tr}} \left\{ \epsilon ^{\mu \nu} ( 
    \partial _{\mu} \omega _{\nu} + \omega _{\mu} \omega _{\nu} ) \phi + 
         \frac {1}{2} \epsilon ^{\mu \nu} B _{\mu \nu} \phi ^2 \right\}, 
                                                               \label {eqn:ba}
\end{eqnarray}
where $\phi$, $\omega_\mu$ and $B_{\mu\nu}$ are 
scalar, vector and antisymmetric tensor fields, respectively, 
and $\epsilon^{01}=1$\footnote{
%%%%%%%%%footnote%%%%%%%%%%%
Throughout this paper we impose 
$\phi^{\dagger} = -\phi, \omega_{\mu}^{\dagger}= - \omega_{\mu}$ 
and $B^{\dagger} = B$ to make the classical action hermitian.
%%%%%%%%%%%%%%%%%%%%%%%%%%%%
}.
This Lagrangian possesses gauge symmetries corresponding to (\ref{eqn:gt})
\begin{eqnarray}
 \delta \phi & = & [ \phi , v_1 ],\label{eqn:dp}\\
 \delta \omega_\mu & = & \partial_\mu v_1 + [ \omega_{1 \mu} , v_1 ]
                          - \{ \phi , u_{1 \mu} \},\label{eqn:do}\\
 \delta B & = & \epsilon^{\mu \nu} ( \partial_\mu u_{1 \nu} + 
                  [ \omega_\mu , u_{1 \nu} ] ) 
                        + [ B , v_1 ] + [ \phi , b_1 ],\label{eqn:db}
\end{eqnarray}
where $B$ is defined 
by $B \equiv \frac {1}{2} \epsilon ^{\mu \nu} B _{\mu \nu}$ 
and $b_1$ by $b_1 \equiv \frac {1}{2} \epsilon^{\mu \nu} b_{1 \mu \nu}$.
Equations of motion of this theory are given by 
\begin{eqnarray}
 \phi : & & - \epsilon ^{\mu \nu} 
                  ( \partial _{\mu} \omega _{\nu} +  
                          \omega _{\mu} \omega _{\nu} )
           - \{ \phi , B \} = 0, \label{eqn:ep} \\ 
 \omega _{\mu} : & & - \epsilon ^{\mu \nu} 
                  ( \partial _{\nu} \phi + [ \omega_{\nu} , \phi ] ) = 0, 
                                                       \label{eqn:eo} \\
 B : & & - \phi ^2 = 0.\label{eqn:eb}
\end{eqnarray}
This system is on-shell reducible since the gauge transformations 
(\ref{eqn:dp})$-$(\ref{eqn:db}) are invariant under the transformations    
\begin{eqnarray}
 \delta v_1 & = & \{ \phi , v_2 \}, \nonumber \\
 \delta u_{1 \mu} & = & \partial_\mu v_2 + [ \omega_\mu , v_2 ]
                                      - [ \phi , u_{2 \mu} ],\nonumber \\
 \delta b_1 & = & \epsilon^{\mu \nu} ( \partial_\mu u_{2 \nu} + 
                  [ \omega_\mu , u_{2 \nu} ] ) 
                        + \{ B , v_2 \} + \{ \phi , b_2 \},\nonumber
\end{eqnarray}
with the on-shell conditions.
However this is not the end of the story.
Indeed this system is infinitely on-shell reducible, {\it i.e.},  
successive reducibilities are given by the following relations; 
\begin{eqnarray}
 \delta v_n & = & [ \phi , v_{n+1} ]_{(-)^{n+1}}, \label{eqn:rv} \\
 \delta u_{n \mu} & = & \partial_\mu v_{n+1} + [ \omega_\mu , v_{n+1} ]
                            - [ \phi , u_{n+1 \mu} ]_{(-)^n},\label{eqn:ru} \\
 \delta b_n & = & \epsilon^{\mu \nu} ( \partial_\mu u_{n+1 \nu} + 
                  [ \omega_\mu , u_{n+1 \nu} ] ) 
              + [ B , v_{n+1} ]_{(-)^{n+1}} + [ \phi , b_{n+1} ]_{(-)^{n+1}},
                                                         \label{eqn:rb} \\
           &&   \hspace{5cm} n = 1,2,3,\cdots, \nonumber
\end{eqnarray}
where $[ \ , \ ]_{(-)^n}$ is a commutator for odd $n$ 
and an anticommutator for even $n$.
This fact is more easily understood by using compact notations 
such as the generalized gauge field $\cal{A}$ and parameter $\cal{V}$.
We define ${\cal{V}}_n$ from $v_n$, $u_{n \mu}$ and $b_n$ by
\begin{eqnarray}
 {\cal{V}}_{2n} & = & {\mbox{\bf j}} u_{2n \mu}dx^{\mu} 
              + {\mbox{\bf k}} \left( v_{2n} + \frac{1}{2} b_{2n \mu \nu} 
                        dx^{\mu} \wedge dx^{\nu} \right) \ \in \Lambda_{-}, \\
 {\cal{V}}_{2n+1} & = & {\mbox{\bf 1}} \left( v_{2n+1} + 
                \frac{1}{2} b_{2n+1 \mu \nu} dx^{\mu} \wedge dx^{\nu} \right)  
                  - {\mbox{\bf i}}  u_{2n+1 \mu}dx^{\mu} \  \in \Lambda_{+}, \\
                & &  \hspace{5cm} n = 0,1,2,\cdots, \nonumber
\end{eqnarray}
where $v_0 = \phi$, $u_{0,\mu} = \omega_{\mu}$ and $b_0 = B$ 
and thus ${\cal{V}}_0 = {\cal{A}}$.
Then eqs.(\ref{eqn:dp})$-$(\ref{eqn:db}) and (\ref{eqn:rv})$-$(\ref{eqn:rb}) 
can be described in the following compact form,
\begin{equation}
  \delta {\cal{V}}_n  = 
             (-)^n [ \ Q + {\cal{A}} \ , \ {\cal{V}}_{n+1} \ ]_{(-)^{n+1}},
                                                        \label{eqn:ir} 
                            \hspace{2cm}   n = 0,1,2,\cdots.
\end{equation}
Using these notations, it is easy to see the on-shell reducibility
\begin{eqnarray}
  \delta {\cal{V}}_n\Bigr\vert_{{\cal V}_{n+1}\rightarrow 
                                   {\cal V}_{n+1}+\delta{\cal V}_{n+1}} 
        & = & (-)^n [ \ Q + {\cal{A}} \ , \ {\cal{V}}_{n+1} 
                   + \delta {\cal{V}}_{n+1} \ ]_{(-)^{n+1}} \nonumber \\
        & = & \delta {\cal{V}}_n + (-)^n \left[ \ Q + {\cal{A}} \ , \ 
                 (-)^{n+1} [ \ Q + {\cal{A}} \ , \ {\cal{V}}_{n+2} \ 
                     ]_{(-)^{n+2}} \ \right]_{(-)^{n+1}}  \nonumber \\
        & = & \delta {\cal{V}}_n - [ \ {\cal{F}} \ , \ {\cal{V}}_{n+2} \ ]
                                                               \nonumber  \\
        & = & \delta {\cal{V}}_n,                     \label{eqn:irr}
\end{eqnarray}
where we used the equation of motion (\ref{eqn:ge}).

Actually the infinite on-shell reducibility is a common feature 
of generalized Chern-Simons theories with nonabelian gauge algebras 
in arbitrary dimensions, which can be understood 
by the fact that (\ref{eqn:irr}) is the relation 
among the generalized gauge fields and parameters.
Thus generalized Chern-Simons theories add another category 
of infinitely reducible
systems to known examples like Brink-Schwarz superparticle~\cite{bs},
Green-Schwarz superstring~\cite{gs}
and covariant string field theories~\cite{w2}.
It should be noted that this theory is infinitely reducible though
it contains only {\it finite} number of fields of {\it finite} rank 
antisymmetric tensors.
Brink-Schwarz superparticle and Green-Schwarz superstring are 
the similar examples in the sense that they contain only finite number 
of fields yet are infinitely reducible.
In the present case the infinite reducibility is understood 
from the following facts:
Firstly, the highest form degrees of ${\cal{V}}_n$ is unchanged from that of
${\cal{V}}_{n-1}$ in eq.(\ref{eqn:ir}) 
since the generalized gauge field $\cal{A}$ contains 
the zero form gauge field $\phi$.
Secondly, the generalized Chern-Simons actions possess the same functional 
form (\ref{eqn:ga}) as the ordinary Chern-Simons action 
and thus have the vanishing curvature condition as the equation of motion;
${\cal{F}} = 0$ (\ref{eqn:ge}).
Thus the equations (\ref{eqn:ir}) representing the infinite reducibilities 
have the same form at any stage $n$, 
except for the difference between commutators and anticommutators. 
Algebraically, the structure of infinite reducibility resembles 
that of string field theories of a Chern-Simons form.

Before closing this section, 
we compare the generalized Chern-Simons theory  of the abelian 
$gl(1,{\bf R})$ algebra, which was investigated previously~\cite{kos}, 
with the model of nonabelian algebra.
In the abelian case commutators in the gauge algebra vanish 
while only anticommutators remain.
Then we can consistently put all transformation parameters to be zero 
except for $v_1$, $u_{1 \mu}$ and $v_2$.
This leads to the previous analysis that the abelian version was quantized 
as a first stage reducible system.
In nonabelian cases, however, infinite reducibility is the universal 
and inevitable feature of the generalized Chern-Simons theories.

%%%%%%%%%%%%%%%%%%%%%%%%%%%%%%%%%%%%%%

\section{Minimal sector} 
\setcounter{equation}{0}

In this section we present a construction of the minimal part 
of quantized action based on the Lagrangian formulation 
given by Batalin and Vilkovisky~\cite{bv}.

In the construction of Batalin and Vilkovisky, ghosts and ghosts for ghosts 
and the corresponding antifields are introduced according to the reducibility 
of the theory.
We denote a minimal set of fields by $\Phi^A$ which include classical fields 
and ghost fields, and the corresponding  antifields by $\Phi_A^{\ast}$.
If a field has ghost number $n$, its antifield has ghost number $-n-1$.
Then a minimal action is obtained by solving the master equation,
\begin{eqnarray}
  (S_{min}(\Phi , \Phi^*),S_{min}(\Phi , \Phi^*)) &=& 0,
                                \label{eqn:me}        \\
  (X,Y)&=&{\partial_r X \over \partial \Phi^A}
          {\partial_l Y \over \partial \Phi_A^*}
          -{\partial_r X \over \partial \Phi_A^*}
          {\partial_l Y \over \partial \Phi^A},
\end{eqnarray}
with the following boundary conditions,
\begin{eqnarray}
  S_{min} \Bigr\vert_{\Phi^{\ast}_A = 0} & = & S_0, \label{eqn:bbc} \\
  \frac{\partial S_{min}}{\partial \Phi^{\ast}_{a_n}} 
                             \Bigr\vert_{\Phi^{\ast}_A = 0} 
           & = & Z^{a_n}_{a_{n+1}} \Phi^{a_{n+1}}, \label{eqn:bbd} 
                \hspace{2cm} n = 0,1,2,\cdots, 
\end{eqnarray}
where $S_0$ is the classical action and $Z^{a_n}_{a_{n+1}} \Phi^{a_{n+1}}$ 
represents the $n$-th reducibility transformation where the reducibility 
parameters are replaced by the corresponding ghost fields.
In this notation, the relation with $n = 0$ in eq.(\ref{eqn:bbd}) 
corresponds to the gauge transformation. 
The BRST transformations of $\Phi^A$ and $\Phi_A^{\ast}$ 
are given by the following equations;
\begin{eqnarray}
 s \Phi^A = (\Phi^A , S_{min}( \Phi , \Phi^{\ast} )), \hspace{1cm}
 s \Phi_A^{\ast} = (\Phi_A^{\ast} , S_{min}( \Phi , \Phi^{\ast} )). 
                                                            \label{eqn:br}
\end{eqnarray}
Eqs.(\ref{eqn:me}) and (\ref{eqn:br}) assure that the BRST transformation 
is nilpotent and the minimal action is invariant under the transformation.
In the present case it is difficult to solve the master equation 
(\ref{eqn:me}) order by order with respect to the ghost number 
because the theory we consider is infinitely reducible.
We need to solve an infinite set of equations according to the introduction of 
an infinite set of ghost fields; ghosts, ghosts for ghosts, 
$\cdots$ and the corresponding antifields.
There is, however, a way to circumvent the difficulties by using 
the characteristics of generalized Chern-Simons theory in which fermionic 
and bosonic fields, and odd and even forms, can be treated in a unified manner.

First we introduce infinite fields
\begin{eqnarray}
 C_n, \ C_{n \mu}, \ \widetilde{C}_n = 
                    \frac{1}{2} \epsilon^{\mu \nu} C_{n \mu \nu}, \hspace{1cm} 
 n = 0, \pm1, \pm2,\cdots,\pm \infty, \label{eqn:cnc}
\end{eqnarray}
where the index $n$ indicates the ghost number of the field.  
The fields with ghost number $0$ are the classical fields
\begin{eqnarray}
 C_0 = \phi, \ \ C_{0 \mu} = \omega_{\mu}, \ \ \widetilde{C}_0 = B. 
\end{eqnarray}
The fields with even (odd) ghost numbers are bosonic (fermionic).
It is seen from eqs.(\ref{eqn:dp})$-$(\ref{eqn:db}) and 
(\ref{eqn:rv})$-$(\ref{eqn:rb}) that fields content for ghosts and 
ghosts for ghosts in the minimal set is completed in the sector for $n>0$ 
while the necessary degrees of freedom for antifields are saturated for $n<0$. 
We will later identify fields with negative ghost numbers as antifields.
We now define a generalized gauge field $\widetilde{{\cal A}}$ 
in such a form of (\ref{eqn:gc}) 
as it contains these infinite fields according to 
their Grassmann parities and form degrees,
\begin{eqnarray}
 \psi & = & \sum_{n = -\infty}^{\infty} C_{2n+1 \mu} dx^{\mu}, \\
 \hat{\psi} & = & \sum_{n = -\infty}^{\infty}  \left( C_{2n+1} + 
             \frac{1}{2} C_{2n+1 \mu \nu} dx^{\mu} \wedge dx^{\nu} \right), \\
 A & = & \sum_{n = -\infty}^{\infty} C_{2n \mu} dx^{\mu}, \\
 \hat{A} & = & \sum_{n = -\infty}^{\infty}  \left( 
          C_{2n} + \frac{1}{2} C_{2n \mu \nu} dx^{\mu} \wedge dx^{\nu} \right).
\end{eqnarray}
We then introduce a generalized action for $\widetilde{\cal A}$ as
\begin{eqnarray}
  \widetilde{S} & = & \frac {1}{2} \int  \ {\mbox{Tr}}^0_{\mbox{\bf {k}}} 
             \left( \widetilde{{\cal A}}Q\widetilde{{\cal A}} 
                + \frac {2}{3} \widetilde{{\cal A}}^3 \right) \label{eqn:ma} \\
  & = & - \int {d^2}x {\mbox{Tr}}^0  
  \left\{
     \sum_{n=-\infty}^{\infty} C_{2n} \left(  \epsilon^{\mu \nu} \partial_{\mu}
         C_{-2n \nu} \raisebox{6mm}{} \right. \right.  \nonumber \\
    &+& \hspace{-0.5cm} \left.\left.
          \sum_{m=-\infty}^{\infty} \left( \epsilon^{\mu \nu}
        C_{2m \mu} C_{-2(m+n) \nu} + \{ C_{2m-1} , \widetilde{C}_{-2(m+n)+1} \}
   - \epsilon^{\mu \nu} C_{2m-1 \mu} C_{-2(m+n)+1 \nu} 
                  \right) \right) \right. \nonumber \\
    &+& \hspace{-0.5cm} \left.  \sum_{n=-\infty}^{\infty} 
                \sum_{m=-\infty}^{\infty}
                       \widetilde{C}_{2n} \left( C_{2m-1} C_{-2(m+n)+1} 
                               + C_{2m} C_{-2(m+n)} \right) \right.\nonumber \\
 &+& \hspace{-0.5cm} \left. 
        \sum_{n=-\infty}^{\infty} C_{2n-1 \mu} \left( \epsilon^{\mu \nu}
               \partial_{\nu} C_{-2n+1} + \sum_{m=-\infty}^{\infty}
                        [ C_{2m \nu} , C_{-2(m+n)+1} ] \right) \right\},
\end{eqnarray}
where the upper index 0 on ${\mbox{Tr}}$ indicates to pick 
up only the part with ghost number 0.
This action is invariant under the following transformation
\begin{eqnarray}
 \delta_{\lambda} \widetilde{{\cal A}} = -\widetilde{{\cal F}} \ {\mbox{\bf i}}
                                                       \lambda,\label{eqn:dl}
\end{eqnarray}
where $\widetilde{{\cal F}}$ is the generalized curvature (\ref{eqn:gf}) 
constructed of $\widetilde{{\cal A}}$ and $\lambda$ is 
a fermionic scalar parameter with ghost number $-1$.
It should be understood that the same ghost number sectors must be equated 
in eq.(\ref{eqn:dl}).
Since $\widetilde{{\cal F}}$ and ${\bf i}\lambda$ belong to $\Lambda_+$
and $\Lambda_-$, respectively, their product in the right hand side 
of eq.(\ref{eqn:dl}) belongs to the same $\Lambda_-$ class as 
$\widetilde{\cal A}$.
The invariance of the action $\widetilde{S}$ under the transformation 
(\ref{eqn:dl}) can be checked by the following manipulation, 
\begin{eqnarray}
 \delta_{\lambda} \widetilde{S} & = & - \int \ {\mbox{Tr}}^0_{\mbox{\bf {k}}} 
        \left\{ ( Q \widetilde{{\cal A}} + \widetilde{{\cal A}}^2 ) 
                     \widetilde{{\cal F}} \ {\mbox{\bf {i}}} \lambda \right\}
                                                    \nonumber \\
 & = &  \int \ {\mbox{Tr}}^0_{\mbox{\bf {j}}} 
                      ( \widetilde{{\cal F}} \widetilde{{\cal F}} ) 
                                                \cdot \lambda \nonumber \\
 & = &  \int \ {\mbox{Tr}}^0_{\mbox{\bf {j}}}
                   \left\{ Q ( \widetilde{{\cal{A}}} Q \widetilde{{\cal{A}}} 
              + \frac{2}{3} \widetilde{{\cal{A}}}^3 ) \right\} \cdot \lambda 
                                                                 \nonumber \\
& = & 0, 
\end{eqnarray}
where the subscript ${\bf j}$ plays the similar role as the subscript 
${\bf k}$, {\it i.e.}, to pick up only the coefficient of ${\bf j}$ 
in the trace.
The change of the subscript ${\bf k}$ to ${\bf j}$ is necessary to take 
${\bf i}$ into account in the trace in accordance with 
${\bf j} {\bf i} = - {\bf k}$.
Here we have simply ignored the boundary term and thus the invariance 
is valid up to the surface term.

We now show that a right variation $s$ defined by 
$\delta_\lambda \widetilde{{\cal A}}=s\widetilde{{\cal A}} \lambda$
is the BRST transformation.
First of all this transformation is nilpotent,
\begin{eqnarray}
 s^2\widetilde{{\cal A}}\lambda_2\lambda_1
 =\delta_{\lambda_2} \delta_{\lambda_1} \widetilde{{\cal A}} =  
        - \delta_{\lambda_2} \widetilde{{\cal F}} \ {\mbox{\bf i}} \lambda_1 = 
            - [ \ Q + \widetilde{{\cal A}} \ , \ \widetilde{{\cal F}} \ ] 
                               \lambda_2 \lambda_1 = 0,
\end{eqnarray}
where the generalized Bianchi identity is used,
\begin{eqnarray}
 [ \ Q + \widetilde{{\cal A}} \ , \ \widetilde{{\cal F}} \ ] = 
    [ \ Q + \widetilde{{\cal A}} \ , \ ( \ Q + \widetilde{{\cal A}} \ )^2 \ ]
        = 0.
\end{eqnarray}
Next we need to show that the transformation $s$ is realized as 
the antibracket form of (\ref{eqn:br}).
The invariance of $\widetilde{S}$ under (\ref{eqn:dl}) implies that 
$\widetilde{S}$ is indeed the minimal action if we make a proper 
identification of fields of negative ghost numbers with antifields.
It is straightforward to see that the BRST transformations 
(\ref{eqn:br}), both for fields and antifields, are realized 
under the following identifications with 
$S_{min}=\widetilde{S}$;
\begin{eqnarray}
\label{eqn:aid}
\begin{array}{rclcrclc}
 C_{-2n+1 \mu} & = & \epsilon_{\mu \nu}^{-1} C_{2(n-1)}^{\nu \ast},& &
 C_{-2n \mu} & = & \epsilon_{\mu \nu}^{-1} C_{2n-1}^{\nu \ast}, & \\
 \raisebox{5mm}{}
 C_{-2n+1} & = & {\widetilde{C}_{2(n-1)}}^{\ast},& &
 C_{-2n} & = & - {\widetilde{C}_{2n-1}}^{\ast}, & \\
 \raisebox{5mm}{}
 \widetilde{C}_{-2n+1} & = & C_{2(n-1)}^{\ast}, & &
 \widetilde{C}_{-2n} & = & - C_{2n-1}^{\ast}, & n = 1,2,3,\cdots,
\end{array}
\end{eqnarray}
where $\epsilon_{\mu \nu}^{-1}$ is the inverse of $\epsilon^{\mu \nu}$, 
$\epsilon^{\mu \rho} \epsilon_{\rho \nu}^{-1} = \delta_{\nu}^{\mu}$\footnote{
%%%%%%%%%%footnote%%%%%%%%%%%%%%
To be precise the antifields are defined as \ 
 $C_n^{\ast} = C^{\ast a}_n \eta_{ab}^{-1} T^b,\cdots,$
with ${\mbox{Tr}} T^{a} T^{b} = \eta^{ab}$. 
%%%%%%%%%%%%%%%%%%%%%%%%%%%%%%%%
}.
This shows that we have obtained a solution for the master equation 
(\ref{eqn:me}), 
\begin{equation}
  \delta_{\lambda} S_{min} 
    = ( S_{min} , S_{min} ) \cdot \lambda = 0.
\end{equation}

It is easy to see that this solution satisfies 
the boundary conditions (\ref{eqn:bbc}) and (\ref{eqn:bbd}),
by comparing the gauge transformation 
(\ref{eqn:dp})$-$(\ref{eqn:db}) and the reducibilities 
(\ref{eqn:rv})$-$(\ref{eqn:rb}) 
with the following expansion of $S_{min}$,
\begin{eqnarray*}
  S_{min} & = &  \int {d^2}x {\mbox{Tr}} \left\{
               - \epsilon ^{\mu \nu} ( 
    \partial _{\mu} \omega _{\nu} + \omega _{\mu} \omega _{\nu} ) \phi - 
         \frac {1}{2} \epsilon ^{\mu \nu} B _{\mu \nu} \phi ^2  \right. \\
     & & \left. + \sum_{n=0}^{\infty} \left\{ \ 
               C_{n}^{\ast} [ \phi , C_{n+1} ]_{(-)^{(n+1)}} 
                                                  \right. \right. \\
     & & \left. \left. \hspace{1cm} + 
                 C_{n}^{\mu \ast} \left( \ \partial_{\mu} C_{n+1} 
                              + [ \omega_{\mu} , C_{n+1} ] 
                      - [ \phi , C_{n+1 \mu} ]_{(-)^n} \ \right)
                                                 \right. \right. \\
     & & \left. \left.\hspace{1cm} + 
               \widetilde{C}_{n}^{\ast} \left( \ \raisebox{4mm}{}
           \epsilon^{\mu \nu} ( \partial_{\mu} C_{n+1 \nu} 
                                   + [ \omega_{\mu} , C_{n+1 \nu} ] ) 
                                         \right. \right. \right. \\
     & & \left. \left. \left.  \hspace{3cm} 
                + [ B , C_{n+1} ]_{(-)^{(n+1)}} 
                + [ \phi , \widetilde{C}_{n+1} ]_{(-)^{(n+1)}} 
                                              \   \right) \ \right\} 
                               + \cdots\cdots \raisebox{5mm}{} \right\}.
\end{eqnarray*}
Thus the action $S_{min}=\widetilde{S}$ 
with the identification (\ref{eqn:aid}) is the correct solution 
of the master equation for the generalized Chern-Simons theory.
It is easy to see that this minimal action also satisfies 
the quantum master equation.

For completeness we give explicit forms of the BRST transformations 
of the minimal fields;
\begin{eqnarray}
 s C_{2n} & = & - \sum_{m = -\infty}^{\infty} 
                                 [ C_{2m+1} , C_{2(n-m)} ], \label{eqn:bc} \\
 s C_{2n-1} & = &  \sum_{m = -\infty}^{\infty} 
                 \left( \frac{1}{2} \{ C_{2m} , C_{2(n-m)} \} + 
                        \frac{1}{2} \{ C_{2m-1} , C_{2(n-m)+1} \} \right),\\
 s C_{2n \mu} & = & \partial_{\mu} C_{2n+1} + \sum_{m = -\infty}^{\infty}
                       \left( [ C_{2m \mu} , C_{2(n-m)+1} ] - 
                             \{ C_{2m+1 \mu} , C_{2(n-m)} \} \right), \\
 s C_{2n-1 \mu} & = & \partial_{\mu} C_{2n} + \sum_{m = -\infty}^{\infty}
                       \left( [ C_{2m \mu} , C_{2(n-m)} ] + 
                             \{ C_{2m-1 \mu} , C_{2(n-m)+1} \} \right), \\
 s \widetilde{C}_{2n} & = & \epsilon^{\mu \nu} \partial_{\mu} C_{2n+1 \nu}
                      + \sum_{m = -\infty}^{\infty} \Big( \epsilon^{\mu \nu} 
                   [ C_{2m \mu} , C_{2(n-m)+1 \nu} ] \nonumber \\ 
       & &    - [ \widetilde{C}_{2m+1} , C_{2(n-m)} ] - 
                     [ C_{2m+1} , \widetilde{C}_{2(n-m)} ] \Big), \\
 s \widetilde{C}_{2n-1} & = & 
                      \epsilon^{\mu \nu} \partial_{\mu} C_{2n \nu} \nonumber \\
              & & +\sum_{m = -\infty}^{\infty} \Big( 
          \frac{1}{2} \epsilon^{\mu \nu} [ C_{2m \mu} , C_{2(n-m) \nu} ] 
                  - \frac{1}{2} \epsilon^{\mu \nu} 
                [ C_{2m-1 \mu} , C_{2(n-m)+1 \nu} ]  \nonumber \\ 
       & &    + \{ C_{2m} , \widetilde{C}_{2(n-m)} \} + 
              \{ C_{2m-1} , \widetilde{C}_{2(n-m)+1} \} \Big), \label{eqn:bd}
\end{eqnarray}
where the identification (\ref{eqn:aid}) should be understood.

It is critical in our construction of the minimal action that the action 
of the generalized theory possesses the same structure as the Chern-Simons 
action and fermionic and bosonic fields are treated in an unified manner. 
It is interesting to note that the starting classical action,
which includes only bosonic fields, and the quantized minimal action, 
which includes the infinite series of bosonic and fermionic fields, 
have the same form of (\ref{eqn:ga}) with the replacement 
${\cal{A}} \rightarrow \widetilde{\cal{A}} $.  
This is reminiscent of string field theories 
whose actions have the Chern-Simons form:
A string field contains infinite series of ghost fields and antifields.
The quantized minimal action also takes the same Chern-Simons form~\cite{w2}.
It is also worth mentioning that there are other examples where
classical fields and ghost fields are treated in a unified way~\cite{ba}.

It is obvious that the minimal action for generalized Chern-Simons theory 
in arbitrary even dimensions can be constructed in the same way 
as in the two-dimensional case  
because the classical action (\ref{eqn:ga}), symmetries (\ref{eqn:gt}), 
reducibilities (\ref{eqn:ir}), the minimal action (\ref{eqn:ma}) and BRST 
transformations 
$ s\widetilde{\cal{A}} = - \widetilde{\cal{F}} {\mbox{\bf{i}}} $ 
are described by using generalized fields and parameters.

%%%%%%%%%%%%%%%%%%%%%%%%%%%%%%%%%%%%%%%%

\section{Gauge fixed action} 
\setcounter{equation}{0}

The gauge degrees of freedom are fixed by introducing a nonminimal action 
which must be added to the minimal one, and choosing a suitable gauge fermion.
Though the number of gauge-fixing conditions is determined in accordance with 
the ``real'' gauge degrees of freedom, 
we can prepare a redundant set of gauge-fixing conditions and then
compensate the redundancy by introducing extraghosts.
Indeed Batalin and Vilkovisky gave a general prescription to construct
a nonminimal sector by this procedure~\cite{bv}.
This prescription is, however,
inconvenient in the present case since it leads to
a doubly infinite number of fields; antighosts, 
extraghosts,$\cdots$, where ``doubly infinite'' means the infinities 
both in the vertical direction and the horizontal direction 
in the triangular tableau of ghosts.      
We can instead adopt gauge-fixing conditions so that 
such extra infinite series do not appear while propagators for all fields 
be well-defined.
The type of gauge-fixing prescription 
which is unconventional for the Batalin-Vilkovisky
 formulation is known, for example, 
in a quantization of topological Yang-Mills theory~\cite{lp}.
In the present case, we found that in each sector of the ghost number
the standard Landau type gauge-fixing for the vector and antisymmetric 
tensor fields is sufficient to make a complete gauge-fixing.  

After taking into account the above points, 
we introduce the following nonminimal action, 
\begin{eqnarray}
 S_{nonmin} = \int d^2 x \sum_{n=1}^{\infty} {\mbox{Tr}} 
                   \left( \bar{C}_{n}^{\ast} b_{n-1}
                    + \bar{C}_{n \mu}^{\ast} b_{n-1}^{\mu}
                    + \eta_{n-1}^{\ast} \pi_{n} \right),
\end{eqnarray}
where the index $n$ indicates a ghost number except that ghost number of 
$b_n$ is $-n$, 
and even (odd) ghost number fields are bosonic (fermionic), as usual.
The BRST transformations of these fields are defined 
by this nonminimal action,
\begin{eqnarray}
  \label{eqn:be}
\begin{array}{rclcrcl}
  s \bar{C}_n & = & b_{n-1},& &
  s b_{n-1} & = & 0,  \\
  s \bar{C}_n^{\mu} & = & b_{n-1}^{\mu},& &
  s b_{n-1}^{\mu} &= &0,\\
  s \eta_{n-1} & = & \pi_{n},& &
  s \pi_{n} & = & 0,\\
  s \bar{C}_n^{\ast} & = & 0,& &
  s b_{n-1}^{\ast} & = &(-)^{n} \bar{C}_n^{\ast}, \\
  s \bar{C}_{n \mu}^{\ast} & = & 0,& &
  s b_{n-1 \mu}^{\ast} & = & (-)^n \bar{C}_{n \mu}^{\ast},\\
  s \eta_{n-1}^{\ast} & = & 0,& &
  s \pi_{n}^{\ast} & = &(-)^{n+1} \eta_{n-1}^{\ast}. 
\end{array}
\end{eqnarray}
Next we adopt the following gauge fermion $\Psi$ 
which leads to a Landau type gauge fixing,
\begin{eqnarray}
 \Psi = \int d^2 x \sum_{n=1}^{\infty} {\mbox{Tr}} \left( 
          \bar{C}_{n} \partial^{\mu} C_{n-1 \mu}
          + \bar{C}_{n}^{\mu} \epsilon_{\mu \nu}^{-1} \partial^{\nu}
                                                       \widetilde{C}_{n-1}
          + \bar{C}_{n}^{\mu} \partial_{\mu} \eta_{n-1} \right), \label{eqn:ps}
\end{eqnarray}
where we assume a flat metric for simplicity. 
Then the antifields can be eliminated by equations 
$\Phi_A^* = \frac{\partial \Psi}{ \ \partial \Phi^A}$,
\begin{eqnarray}
 C_n^{\ast} & = & 0,\label{eqn:ab}\\
 C_n^{\mu \ast} & = & - \partial^{\mu} \bar{C}_{n+1},\\
 \widetilde{C}_n^{\ast} & = & \epsilon_{\mu \nu}^{-1} \partial^{\mu} 
                                                   \bar{C}_{n+1}^{\nu},\\
 \bar{C}_{n+1}^{\ast} & = & \partial^{\mu} C_{n \mu},\\
 \bar{C}_{n+1 \mu}^{\ast} & = & \epsilon_{\mu \nu}^{-1} \partial^{\nu} 
                                  \widetilde{C}_n + \partial_{\mu} \eta_n,\\
 \eta_{n-1}^{\ast} & = & - \partial_{\mu} \bar{C}_n^{\mu},\label{eqn:ae}
                     \hspace{3cm} n = 0,1,2,\cdots.
\end{eqnarray}
The complete gauge-fixed action $S_{tot}$ is 
\begin{eqnarray}
 S_{tot} = S_{min}|_{\Sigma} + S_{nonmin}|_{\Sigma}, 
\end{eqnarray}
where $\Sigma$ is a surface defined by 
eqs.(\ref{eqn:ab})$-$(\ref{eqn:ae}).
This action is invariant under the on-shell nilpotent BRST transformations
(\ref{eqn:bc})$-$(\ref{eqn:bd}) and (\ref{eqn:be}) 
in which the antifields are eliminated by substituting 
eqs.(\ref{eqn:ab})$-$(\ref{eqn:ae}).
It can be seen that the propagators of all fields are well-defined, 
by writing the kinetic terms and the gauge-fixing terms in $S_{tot}$, 
\begin{eqnarray*}
  S_{tot} & = & \int d^2 x {\mbox{Tr}} \Big\{ \ 
               - \phi \epsilon^{\mu \nu} \partial_{\mu} \omega_{\nu} 
               +  \partial^{\mu} \omega_{\mu} b_0 + 
               \epsilon_{\mu \nu}^{-1} \partial^{\nu} B \ b_0^{\mu}   \\
          & &  \ \ \ \ \ \ \ \ \ \ \ \  \ \  
                + \sum_{n=1}^{\infty} \Big( 
                  - \partial^{\mu} \bar{C}_n \partial_{\mu} C_n
                  - \frac{1}{2} ( \partial^{\mu} \bar{C}_n^{\nu} 
                                 - \partial^{\nu} \bar{C}_n^{\mu} ) 
                                ( \partial_{\mu} C_{n \nu} 
                                 - \partial_{\nu} C_{n \mu} ) \Big)  \\
        & & \ \ \ \ \ \ \ \ \ \ \ \ \ \ 
               +  \sum_{n=1}^{\infty} \left( 
                \partial^{\mu} C_{n \mu} b_{n} + 
                \epsilon_{\mu \nu}^{-1} \partial^{\nu} 
                        \widetilde{C}_{n} b_n^{\mu}
                      + \partial_{\mu} \eta_{n-1}  b_{n-1}^{\mu} - 
               \partial_{\mu} \bar{C}_n^{\mu} \pi_n \right) \\
         & & \ \ \ \ \ \ \ \ \ \ \ \ \ \ 
                      + \ \mbox{ interaction \ terms } \ \Big\}.
\end{eqnarray*}
Thus the gauge fermion (\ref{eqn:ps}) is a correct choice
and the gauge degrees of freedom are fixed completely.
We can consistently determine the hermiticity of the fields with a convention 
$\lambda^{\dagger} = - \lambda$ in eq.(\ref{eqn:dl})\footnote{
%%%%%%%%%%%footnote%%%%%%%%%%%
Hermiticity conditions;
\begin{eqnarray*}
  C_n^{\dagger} & = & - C_n, \ \ 
  C_{n \mu}^{\dagger} = (-)^{n+1} C_{n \mu}, \ \ 
  \widetilde{C}_n^{\dagger} = \widetilde{C}_n, \ \ 
  \bar{C}_n^{\dagger} = (-)^{n+1} \bar{C}_n, \ \ 
  \bar{C}_n^{\mu \dagger} = - \bar{C}_n^{\mu}, \\
  b_n^{\dagger} & = & - b_n, \ \ b_n^{\mu \dagger} = (-)^n b_n^{\mu}, \ \ 
  \eta_n^{\dagger} = \eta_n, \ \ \pi_n^{\dagger} = (-)^{n+1} \pi_n.
\end{eqnarray*}
%%%%%%%%%%%%%%%%%%%%%%%%%%%%%%
}.

Here comes a possible important comment.
There is a universal feature for models of infinitely reducible system 
with finite degrees of freedom, 
that the number of the ``real'' gauge degrees of freedom 
is the half of the original degrees of freedom~\cite{bs,gs}.
The known examples of infinitely reducible system have the same 
characteristics. 
In the present two-dimensional model, there are four parameters 
$v_{n}, u_{n \mu}$ and $b_{n}$ for each stage of the reducibility.
The ``real'' number of gauge-fixing conditions is $3-1=2$,
where three gauge fixing conditions 
${\partial}^{\mu} C_{n-1 \mu} = 0$,
$ {\epsilon}^{-1}_{\mu \nu} {\partial}^{\nu} {\widetilde{C}}_{n-1} = 0 $ 
are linearly dependent due to 
$ {\partial}^{\mu} ( {\epsilon}^{-1}_{\mu \nu} 
                    {\partial}^{\nu} {\widetilde{C}}_{n-1} ) = 0 $ 
and thus we needed to impose an extra condition 
$ {\partial}_{\mu} {\bar{C}}^{\mu}_{n} = 0 $.

%%%%%%%%%%%%%%%%%%%%%%%%%%%%%%%%%%%%%%

\section{Conclusions and discussions} 
\setcounter{equation}{0}

We have investigated the quantization of two-dimensional version of
the generalized Chern-Simons theory with a nonabelian gauge algebra 
by the Lagrangian formalism~\cite{bv}. 
We have found that models formulated by the generalized Chern-Simons theory 
are in general infinitely reducible 
and thus the quantization is highly nontrivial. 
We have derived the on-shell nilpotent BRST transformation 
and the BRST invariant gauge-fixed action for this infinitely reducible system.
We have confirmed that the propagators of all fields are well-defined 
in the gauge-fixed action.  
It is important to recognize that 
the starting classical action includes only bosonic fields, 
while the quantized minimal action includes infinite series of both 
bosonic and fermionic ghost fields, which are treated 
in a unified way by the generalized Chern-Simons formulation.
It is a characteristic of the generalized Chern-Simons theory that 
the quantized minimal action has the same Chern-Simons form as 
the classical action.

The quantization is successfully carried out while there appear 
other possible problems in connection with the introduction of 
the infinitely many fields.
It is then an important question whether we can treat 
the quantum effects of the infinitely many ghost fields consistently.
We have obtained some evidences that quantum effects of the infinitely 
many ghost fields can be treated in a systematic way and lead to a finite 
contribution.
To be specific as a related example, the classical action is independent 
of the space-time metric, but it is not obvious 
that the quantized theory is topological because of the on-shell reducibility.
The similar situation occurs in the nonabelian BF theories~\cite{bbrt}.
We can, however, prove the metric independence of the partition 
function by regularizing the quantum effects of infinitely many ghosts 
contributions in a specific but natural way.
It is also important to analyze quantum effects of correlation 
functions for physical operators.
The details of these points will be given 
in a subsequent publication~\cite{kstu}.

It is interesting to consider physical aspects of an introduction 
of the infinite number of ghost fields. 
An immediate consequence is a democracy of ghosts and classical fields, 
{\it i.e.}, the classical fields are simply the zero ghost number sector 
among infinitely many ghost fields. 
The classical gauge fields and ghost fields have no essential difference 
in the quantized minimal action.
In the present paper we have not introduced fermionic gauge fields 
in the starting action but it is straightforward to introduce 
fermionic gauge fields~\cite{kw1} and carry out quantization. 
The classical fermionic fields are just zero ghost number sector 
among infinitely many ghost fields in a quantized action, 
just the same as in the bosonic sector.  
It is tempting to speculate that fermionic matter fields may be identified 
as a special and possibly infinite combination of ghost fields 
because the fermionic and bosonic sectors couple 
in the standard covariant form in the quantized minimal action of 
the generalized Chern-Simons theory.

In the analyses of the quantization of the generalized Chern-Simons theory 
with abelian $gl(1,{\bf R})$ algebra, 
it was pointed out that a physical degree of freedom which 
did not exist at the classical level 
appeared in the constant part of the zero form field $\phi$ 
at the quantum level due to the violation of the regularity~\cite{kos}. 
We know that a zero form field plays an important role in the generalized 
Chern-Simons theories as emphasized in the classical discussion~\cite{kw2,kw3}.
In particular a constant component of the zero form field played 
a role of physical order parameter between the gravity and nongravity phases.
We find it is important to clarify the mechanism how the physical 
constant mode of the zero form field plays the role of possible 
order parameter in the quantum level.
This question is essentially 
related to the regularity violation in the nonabelian version of 
the generalized Chern-Simons theory.
It is, however, expected that this question will be better clarified  
in the Hamiltonian formalism quantization. 
We have already found that the BRST invariant gauge-fixed action obtained from 
the Hamiltonian formalism coincides with that of the Lagrangian formulation. 
These points will also be discussed in a subsequent publication~\cite{kstu}.

Finally we point out that the quantization procedures of the 
generalized Chern-Simons theories given in this paper is universal 
and thus naturally extended to arbitrary even dimensions.
To derive nonminimal action, however, we need to count the genuine 
independent degrees of freedom in the gauge transformation 
and impose a gauge-fixing by choosing an adequate gauge fermion. 
It seems to be a general feature that the independent gauge degrees 
of freedom is just a half of the original degrees of freedom. 
In the Hamiltonian formalism we found a reasoning 
that this should be the case.
 
%%%%%%%%%%%%%%%%%%%%%%%%%%%%%%%%%

\vskip 1cm

\noindent{\Large{\bf Acknowledgments}}\\
One of the authors (N.K.) wishes to thank M.A. Vasiliev for useful comments.
The work by N.K. and K.S. is supported in part by
the Grant-in-Aid for Scientific Research from the Ministry of Education,
Science and Culture (No. 07044048).
One of authors (H.U.) is partially supported by Nukazawa Science Foundation.

\end{document}